\documentclass{aa}
\usepackage{graphicx}

\begin{document}

\title{Quiescent and flare analysis for the chromospherically active 
star \object{Gl\,355} (\object{LQ\,Hya})}

\author{S.~Covino\inst{1} \and 
	M.R.~Panzera\inst{1} \and
	G.~Tagliaferri\inst{1} \and
	R.~Pallavicini\inst{2}} 

\offprints{S. Covino, \email{covino@merate.mi.astro.it}}

\institute{Brera Astronomical Observatory, Via Bianchi 46, I-23807 Merate, 
Italy
\and Palermo Astronomical Observatory, P.za del Parlamento 1, I-90134 Palermo, 
Italy}

\date{Received date / accepted date}

\abstract{
We discuss ROSAT and ASCA observations of the young active star 
\object{Gl\,355}. During the ROSAT observation a strong flare was detected 
with 
a peak flux more than an order of magnitude larger than the quiescent level.  
Spectral analysis of the data allows us to study the temperature and emission 
measure distribution, and the coronal metal abundance, for the quiescent phase 
and, in the case of ROSAT, also during the evolution of the flare. The global 
coronal metallicity $Z/Z_{\odot} \sim 0.1$ derived from both ROSAT and ASCA 
data 
is much lower than solar and presumably also much lower than the photospheric 
abundance expected for this very young star. The temperature structure of the 
quiescent corona was about the same during the various observations, with a 
cooler component at $T_1 \sim  7$ MK and a hotter component (to which only 
ASCA 
was sensitive) at $T_2 \sim 20$ MK. During the flare, the low temperature 
component remained approximately constant and equal to the quiescent value, 
while the high-temperature component was the only one that varied. We have 
modeled the flare with the hydrodynamic-decay sustained-heating approach of 
Reale at al. (1997) and we have derived a loop semi--length of the order of 
$\sim 1.5$ stellar radii, i.e. much larger than the dimensions of flares on 
the 
Sun, but comparable with the typical dimensions inferred for other stellar 
flares. We have compared the derived loop size with that estimated with a 
simpler (but physically inconsistent) approach, finding that for this, as well 
for several other stellar flares, the two methods give comparable loop sizes. 
Possible causes and consequences of this result are discussed.
}

\titlerunning{Flare analysis for \object{Gl\,355}}
\authorrunning{Covino et al.\ }

\maketitle

\section{Introduction}

It has been known for a long time that a subset of late K and M dwarf stars 
in the solar neighborhood are characterized by the occurrence of flares at 
optical, UV, X--ray and radio wavelengths. Moreover, all types of late--type 
stars show flaring activity in their soft X--ray emission (see for instance 
Schmitt \cite{Sch94}). Flares on the \object{Sun} and on dKe--dMe stars are 
usually believed to be basically similar in their origin and development,
in spite of the fact that stellar flares are normally $2 \div 4$ orders of 
magnitude more energetic than the largest solar flares. Magnetic coupling 
between the components in a binary system or between a young star and an 
accretion disk has been invoked to explain the energy budget for some giant 
flares (Graffagnino et al. \cite{GWS95}, Grosso et al. \cite{GMF97}, van den 
Oord \cite{v88}). 

One of the fundamental problems in the study of stellar flares is the 
determination of the geometry and size of the flaring coronal plasma. Indeed 
stellar coronae, as shown by spatially resolved observations of the solar 
corona, are far from being spatially homogeneous. The lack of spatial 
information is a strong limitation for the study of stellar coronae:
low--resolution coronal X--ray spectra are insensitive to the plasma density
and do not allow distinguishing between an extended low-pressure emitting 
region 
and a compact high pressure one. Indeed, spatial information about the size of 
a 
flare could help to discriminate between different theoretical scenarios. 
For instance, it is accepted that the decay time of a flare X--ray light curve 
is related to the length of the flaring loop. This property is often used to 
derive the spatial size of unresolved flares, assuming that there is no
heating in the decay (e.g. Haisch \cite{Hai83}, White et al. \cite{WCP86}, van 
den Oord \& Mewe \cite{vM89}, Pallavicini et al. \cite{PTS90}, Pallavicini 
\cite{P95b}). However, if a significant amount of heating is present during 
the 
flare decay, the derived loop lengths can be in error.

Thus far, the main tools to study the spatial distribution of stellar coronal 
plasmas have been eclipse monitorings and the study of flares. However, as 
discussed by Schmitt (\cite{S98}) convincing examples of rotationally 
modulated X--ray emission are rare and eclipse observations often do not 
produce 
unambiguous results (see for instance White et al. \cite{WSP90}, Culhane et 
al. 
\cite{CWSP90}, Tagliaferri et al. \cite{TWD91}, Schmitt \& K\"urster 
\cite{SK93}, Ottmann et al. \cite{OSK93}, K\"urster \& Dennerl \cite{KD93}, 
White et al. \cite{WNA94}, Ottmann \cite{O94}, Antunes et al. \cite{ANW94}, 
Huenemoerder \cite{H98}, Tagliaferri et al. \cite{TCCP99}, Rodon\`o et 
al. \cite{RPL99}). Recently the observation of the total eclipse of a large 
flare on \object{Algol} (Schmitt \& Favata \cite{SF99}) has for the first time 
yielded a strong geometrical constraint on the size of a flaring structure. 
However, apart from this unique case, most estimates of stellar flare sizes 
have 
relied on modeling approaches (e.g. Ortolani et al. \cite{OPMRW98}, Favata and 
Schmitt \cite{FS99}, Maggio et al. \cite{MPRT00}).

Several reviews of the X--ray properties of flare stars based on observations 
performed by different satellites and/or detectors have appeared in the
literature: from {\it Einstein} data (Haisch \cite{Hai83}, Ambruster et al. 
\cite{ASG89}); EXOSAT data (Pallavicini et al. \cite{PTS90}); and ROSAT data 
(Schmitt \cite{Sch94}). A review about the status of solar and stellar flare 
research has been published by Haisch et al. (\cite{HSR91}) and see also 
Haisch 
\& Rodon\`o (\cite{HR89}).

We present here the analysis of an intense X-ray flare detected from 
\object{Gl\,355} with the ROSAT satellite. The high count--rate has allowed us 
to perform a time--resolved spectral analysis of the flare to discuss the 
temporal evolution of plasma parameters such as the temperature $T$, the 
emission measure $EM$, the global coronal metallicity $Z$, and of the 
absorbing 
column density $N_{\rm H}$. We have investigated the possible variation during 
the flare of the metal abundance and of the hydrogen column density. 
The flare analysis was performed by the method developed by Reale et al. 
(\cite{RBPSM97}) which considers the possibility of sustained heating during 
the 
flare decay and the derived loop size was compared with that computed 
following 
the methodology of Pallavicini (\cite{P95b}). The same comparison was 
performed for flares detected from other stars and studied with the Reale et 
al. 
(\cite{RBPSM97}) methodology. The quiescent emission from \object{Gl\,355} was 
also studied using ROSAT and ASCA observations and the long--term behavior of 
\object{Gl\,355} was investigated comparing the results of observations 
performed at different epochs.

The paper is organized as follows. In Section\,\ref{sec:target} we describe 
the 
main parameters and previous observations for \object{Gl\,355}. The ROSAT and 
ASCA observations (light curves and spectra) are discussed in 
Section\,\ref{sec:observations}. In Section\,\ref{sec:quie} we discuss the 
results for the quiescent emission, while the flare analysis and modeling are 
reported in Section\,\ref{sec:flare} and \ref{sec:model}. Finally, in 
Section\,\ref{sec:concl} we present a summary of the main results and draw the 
conclusions.

\section{The target}
\label{sec:target}

\begin{table*}[ht]
\begin{center}
\begin{tabular}{|cc|cc|}
\hline
$\alpha_{2000} = 09^h\,32^m\,25.5^s$ &
$\delta_{2000} = -11^\circ\,11'\,05"$ &
$l_{II} = 244.59$ & $b_{II} = 28.40$ \\
$V = 7.82 \pm 0.02$ & $B-V = 0.93 \pm 0.02$ &
$\pi = 54.52 \pm 0.99$\,mas & $d = 18.34 \pm 0.33$\,pc \\
$V_{rad} = 8.6 \pm 0.5$\,km\,s$^{-1}$ & Spectral Type = K2Ve & 
$P_{rot} = 1.60$\,days & B.C. = $-0.40 \pm 0.05$ \\
$v \sin i = 26.5 \pm 0.5$\,km\,s$^{-1}$ & $i = 55 \pm 5^\circ$ &
W(Li) = 219\,m\AA & $R = 0.8 \pm 0.1\,R_\odot$ \\
\hline
\end{tabular}
\end{center}
\caption{\object{Gl\,355} main parameters.}
\label{tab:HTC}
\end{table*}

\object{Gl\,355} (\object{HD\,82558}, \object{LQ\,Hya}, \object{BD\,-10\,2857},
\object{FK\,S9098}, \object{SAO\,155272}) is a relatively well known nearby 
star of spectral type K2Ve. Some parameters derived from the Hipparcos Input 
Catalogue (HIC, see Turon et al. \cite{TuronHIC92}) and from the 
Hipparcos/Tycho Catalogues HTC, ESA \cite{ESA97}), are reported in 
Table~\ref{tab:HTC}. The rotational period (P$_{rot} = 1.60$ days) and 
the Bolometric Correction (BC) are quoted from Strassmeier et al. 
(\cite{SBCR97}), Fekel et al. (\cite{FMH86}), Robinson et al. (\cite{RCSNS94}) 
and Kurucz (\cite{K93}). $v \sin i$ and the inclination angle $i$ 
are quoted from Donati (\cite{D99}) whereas the lithium line equivalent
width W(Li) is from Sterzik \& Schmitt (\cite{SS97}). The spectral type is 
from 
Montes et al. (\cite{MSCU99}). \object{Gl\,355} has also been classified as a 
BY\,Dra 
variable (Fekel et al. \cite{FBA86} and \cite{FMH86}). Its high lithium 
abundance and high rotation rate suggest a young age, possibly even a 
pre--main sequence object (Vilhu et al. \cite{VGW91}) or more likely a
ZAMS star. Variable spot distributions on this star inferred from Doppler 
imaging have been reported (Saar et al. \cite{SPT92} and \cite{SPT94}, 
Strassmeier et al. \cite{SRWHM93}, Rice \& Strassmeier \cite{RS98}) and 
widespread magnetic fields have been detected (Saar et al. \cite{SPT92} and 
\cite{SPT94}, Basri \& Marcy \cite{BM94}, Donati et al. \cite{DSCRC97}, Donati 
\cite{D99}). \object{Gl\,355} shows chromospheric emission in lines such as 
Ca\,II H and K (Fekel et al. \cite{FBA86} and \cite{FMH86}, Strassmeier et al. 
\cite{SFBDH90}), Ca\,II $\lambda$8542 (Basri \& Marcy \cite{BM94}) and 
$H\alpha$ 
(Vilhu et al. \cite{VGW91}). Strong UltraViolet (UV) chromospheric and 
transition region emission lines were also found by Simon \& Fekel 
(\cite{SF87}). \object{Gl\,355} was also monitored photometrically since 1982 
(Strassmeier et al. \cite{SBCR97}) and two periods were singled out in the 
light--curve: 11.4 and 6.8 years (Ol\`ah et al. \cite{OKS00}). There are also 
strong indications for a short (few years) magnetic cycle (Kitchatinov et al. 
\cite{KJD00}).

The star was detected in the R\"ontgensatellit (ROSAT) All Sky Survey (RASS, 
Snowden \& Schmitt \cite{SS90}) where it is indicated as object 
{1RXS\,J093225.5--111101}. The X--ray data were discussed by Hempelmann et al. 
(\cite{HSSRS95}). The RASS count rate was $2.73\pm0.20$ ct s$^{-1}$ 
and the hardness ratio, $HR$, was $-0.04\pm0.07$. The conversion factor from 
count--rate to flux (in the 0.1--2.4 keV energy band) is $(8.31+5.30 \cdot HR) 
\times 10^{-12}$\,erg\,cm$^{-2}$\,ct$^{-1}$ (Fleming et al. \cite{FMMW94}), 
where HR is the hardness ratio. The RASS flux is therefore $2.2(\pm0.3) \times 
10^{-11}$\,erg\,cm$^{-2}$\,s$^{-1}$. This translates into $\log L_{\rm X} / 
L_{\rm bol} = -3.06$ (where $f_{\rm bol} = L_{bol} / 4 \pi d^2 = 2.7 \times 
10^{-5} \times 10^{-0.4(m_{\rm V} + BC)}$\,erg\,cm$^{-2}$\,s$^{-1}$) putting 
this star toward the highest activity limits (see Sterzik \& Schmitt 
\cite{SS97} 
for a discussion)  very likely in the so--called saturation regime (Randich 
\cite{R00}). With the adopted distance (Table~\ref{tab:HTC}) the X-ray 
luminosity in the RASS observation turns out to be $8.8 \times 
10^{29}$\,erg\,s$^{-1}$ (see also the RASS catalogue of nearby stars of 
H\"unsch 
et al. \cite{HSSV99}). 

\object{Gl\,355} is also among the sources detected in the ROSAT Wide Field 
Camera (WFC) All-Sky Survey (Pye et al. \cite{PMABBDPRSW95}) of 
extreme--ultraviolet sources (\object{RE\,J0932--111}) with a count rate of 
$42\pm6$ and $45\pm7$\,ct\,ks$^{-1}$ in the S1 and S2 bands ($90 \div 206$ and 
$62 \div 110$\,eV), respectively. It has also been observed by the Extreme 
UltraViolet Explorer (EUVE) and results are reported in the all--sky catalogue 
of faint extreme UV sources (Lampton et al. \cite{LLS97}, 
\object{EUVE\,J0933--111}, count--rate $0.066$\,ct s$^{-1}$), in the second 
source catalog (Bowyer et al. \cite{BLLWJM96}, \object{2EUVE\,J0932--11.1}, 
count--rate $77\pm13$\,ct\,ks$^{-1}$ at 100\,\AA ) and spectral atlas (Craig 
et 
al. \cite{CAF97}).

Finally, strong flares in the UV from \object{Gl\,355} have been detected by 
Ambruster \& Fekel (\cite{AF90}) and Montes et al. (\cite{MSCU99}) with the 
International Ultraviolet Explorer (IUE) satellite. However, no flare in 
X--rays 
had previously been detected.

\section{Observations and analysis}
\label{sec:observations}

The observations discussed in this paper were performed by the ROSAT and ASCA 
satellites in Nov 1992 and May 1993, respectively.

A pointed observation of \object{Gl\,355} was performed in Nov 1992 with the 
Position 
Sensitive Proportional Counter (PSPC) detector on board the ROSAT satellite 
(Tr\"umper \cite{Tr83}, Pfeffermann et al. \cite{Pfe87}). The PSPC has an 
energy 
resolution $(\Delta E /E)$ of $\simeq 0.42$ at 1\,keV and a bandwidth of $0.1 
\div 2.4$\,keV. The spectral resolution is quite moderate when compared with 
that of the ASCA detectors but at the lower energies the PSPC is more 
sensitive 
to the presence of soft emission components. 
 
The Advanced Satellite for Cosmology and Astrophysics (ASCA, Tanaka et al. 
\cite{TIH94}) is an X--ray observatory carrying four detectors onboard, namely 
two Solid State Imaging Spectrometers (SIS, Burke et al. \cite{BMH91}) and two 
Gas Imaging Spectrometers (GIS, Ohashi et al. \cite{OMI91}). Each detector
is at the focus of an imaging thin foil grazing incidence telescope and each 
SIS has four CCD chips, but for the observation of \object{Gl\,355} in May 
1993 
they were 
operated in a 1--CCD mode, that implies a Field of View (FoV) of $11' \times 
11'$. The Full Width at Half Maximum (FWHM) energy resolution of each SIS is 
$\sim 60-120$\,eV from 1--6\,keV, compared to 200--600\,eV for the GIS. Each 
GIS has a 40' diameter circular FoV. SIS0 and GIS2 are the two best calibrated 
detectors. The energy bandwidth with $>10$\% efficiency is $0.5 \div 10$\,keV 
for the SIS and $0.8 \div 10$\,keV for the GIS. Screening of the data were 
applied removing data acquired during satellite passages through regions with 
geomagnetic rigidity $< 6$\,GeV/c for the SIS and $< 7$\,GeV/c for the GIS 
(Day 
et al. \cite{DAE95}). 

Light curves and spectra were extracted using the FTOOLS (v. 4.0) package.
The light curve analysis was performed with the XRONOS (v. 4.02) package
while for the spectral analysis we used the XSPEC (v. 10.0) package. Auxiliary
response files were computed with the FTOOLS utilities {\it ascaarf} and {\it 
pcaarf}. Images were analyzed with the XIMAGE package (v. 2.60).

\subsection{ROSAT light curve}
\label{sec:rpsps}

\begin{table*}[ht]
\begin{center}
\begin{tabular}{|ccccc|}
\hline
ID                 & Count rate    	& Soft Band 		& Hard band  	
    & Hardness Ratio \\
                   & $0.1\div2.4$\,keV 	& $0.1\div0.28$\,keV    & 
$0.5\div2.0$\,keV &  (H-S)/(H+S)   \\
\hline
RP200998N00 & $1.73\pm0.04$ 	& $0.64\pm0.02$  	& $0.94\pm0.03$     & 
$0.19\pm0.04$  \\
RP200999N00 & $29.9\pm0.14$ 	& $9.00\pm0.08$ 	& $19.1\pm0.11$     & 
$0.36\pm0.03$  \\
RP201000N00 & $12.8\pm0.09$ 	& $3.54\pm0.05$		& $8.54\pm0.08$     & 
$0.41\pm0.04$  \\
RP201001N00 & $9.11\pm0.08$ 	& $2.70\pm0.04$ 	& $5.80\pm0.06$     & 
$0.36\pm0.03$  \\
RP201002N00 & $4.73\pm0.05$ 	& $1.59\pm0.03$   	& $2.87\pm0.04$     & 
$0.29\pm0.03$  \\
RP201003N00 & $3.02\pm0.04$ 	& $1.11\pm0.03$  	& $1.69\pm0.03$     & 
$0.21\pm0.03$  \\
RP201004N00 & $3.87\pm0.05$ 	& $1.39\pm0.03$  	& $2.14\pm0.04$     & 
$0.21\pm0.03$  \\
RP201005N00 & $3.26\pm0.05$ 	& $1.13\pm0.03$  	& $1.88\pm0.03$     & 
$0.25\pm0.01$  \\
\hline
\end{tabular}
\end{center}
\caption{Parameters computed for the ROSAT PSPC pointed observation of 
\object{Gl\,355}.
Column\,(1): the observation ID; Column\,(2): the background subtracted source 
count--rate in counts\,s$^{-1}$ corrected for the telescope vignetting and 
point spread function; Column\,(3) and (4): the soft and hard band count rates
in counts\,s$^{-1}$, also corrected for the telescope vignetting and point 
spread function; Column\,(5): the computed hardness ratio defined as 
(H-S)/(H+S) where H and S are the hard and soft counts, respectively.}
\label{tab:rosatobspar}
\end{table*}

The ROSAT pointed observation consists of eight shots with exposure times in
the range $\sim 1500 \div 2000$\,s. They were obtained from 1992, November 5 
at 
15:40 Universal Time (UT) to November 6 at 3:35 UT. We retrieved the raw data
(identification codes: RP200998N00--RP201005N00) from the ROSAT public archive.
These observations were also included in the White, Giommi \& Angelini catalog 
(WGA, White et al. \cite{WGA94}, source \object{1WGA\,J0932.4--1110}) where 
parameters as count--rates, hardness  and softness ratios, etc., were 
automatically derived. However, the WGA energy bands are not always suitable 
for 
stellar corona analysis (see for instance Fleming et al. \cite{FMMW94}) and 
therefore we have computed the relevant parameters in more adequate energy 
bands. These are the ``total'' band $\equiv$ PSPC  channels 11--240 ($\approx$ 
0.1--2.4\,keV); ``soft'' band $\equiv$ PSPC channels 11--41 ($\approx$ 
0.1--0.28\,keV); and ``hard'' band $\equiv$ PSPC channels 52--201 ($\approx$ 
0.5--2.0\,keV). Results are reported in Table\,\ref{tab:rosatobspar}. 

\begin{figure}
\resizebox{\hsize}{!}{\includegraphics{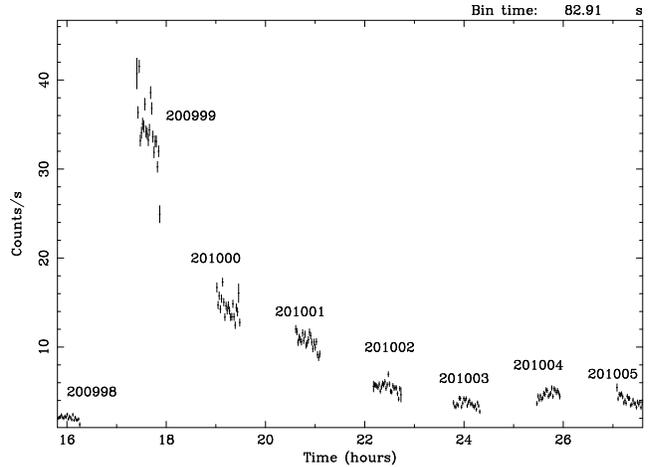}}
\caption{Total light curve obtained by the ROSAT PSPC observations. The count
rate is not background subtracted. Observations were performed starting on 
1992, 
November 5.}
\label{fig:totlc}
\end{figure}

The ROSAT PSPC light curve shows an evident flare which occurred on 1992, Nov 
5, 
at $\sim 18$ UT. The count rate, as reported in Table\,\ref{tab:rosatobspar}, 
increases by more than an order of magnitude. Fig.\,\ref{fig:totlc} shows the 
complete light curve with superimposed the identification codes of the  
observations considered here. The flare maximum can be located close to the 
observation 200999 or just before it. A second, much smaller, event is located 
around the observation 201004 where an increase in the count rate is again 
recorded. The background estimated in circles around the source varies among 
the 
various pointings from $\sim 0.3$\,ct\,s$^{-1}$ to $\sim 0.6$\,ct\,s$^{-1}$ 
and 
it amounts at most to $\sim 20$\% of the source counts.

\begin{figure}
\resizebox{\hsize}{!}{\includegraphics{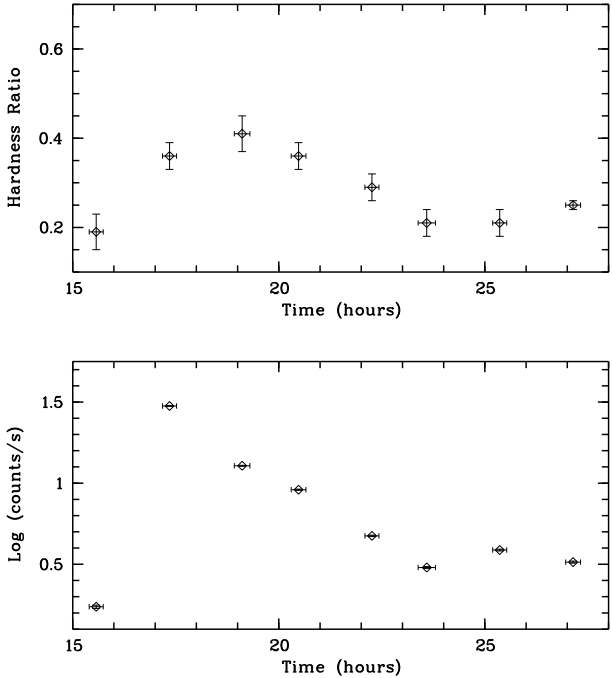}}
\caption{Hardness ratio and background subtracted light curve (in the 
0.1--2.4\,keV band) for the ROSAT PSPC observation. The hardness ratio is 
defined as H-S/H+S where H are the counts in the band between 
0.1--0.28\,keV and S the counts in the band between 0.5--2.0\,keV. The 
resulting 
e--folding time of the light curve is $\tau_{\rm lc} = 10.1 \pm 0.5$\,ks. For 
comparison, the hardness ratio measured during the RASS observations is $-0.04 
\pm 0.07$.}
\label{fig:LC}
\end{figure}

Fig.\,\ref{fig:LC} shows the hardness ratio and background subtracted light 
curve for all the ROSAT pointings. Surprisingly, the hardest observation is 
that 
correspondent to the pointing number 201000, already in the flare decay phase. 
This issue will be discussed later in the context of the flare spectral 
analysis 
(Sect.\,\ref{sec:flare}). The light curve after the flare maximum can be well 
fitted by an exponential with e--folding decay time of $\tau_d \sim 
10.1$\,ks. 

The RASS ROSAT observation performed during the first half of 1990 allows a 
comparison with the pre--flare (200998) and post--flare (201003, 201004, and 
201005) pointings. The count rate of the RASS observation is $\sim 50\%$ 
higher 
than during the pre--flare but almost a factor 2 lower than during the last 
three observations whereas the hardness ratio ($-0.04 \pm 0.07$) seems to 
indicate a softer status than those observed in November 1992. Intrinsic 
long--term variability both in count--rate and spectral shape thus emerges 
(see 
also the ASCA light curve, Sect.\,\ref{sec:aobs}, and the spectral analysis of 
the quiescent emission, Sect.\,\ref{sec:quie}).

\subsection{ASCA light curve}
\label{sec:aobs}

\object{Gl\,355} was observed by the ASCA satellite on 1993, May 7 at 20:14, 
roughly six month after the ROSAT observation. The effective exposures were 
19.1 and 22.2\,ks long for the SIS and GIS detectors, respectively. The 
sequence number of the observation is 21020000 and again we retrieved the raw 
data from the ASCA public archive. 

\begin{figure}
\resizebox{\hsize}{!}{\includegraphics{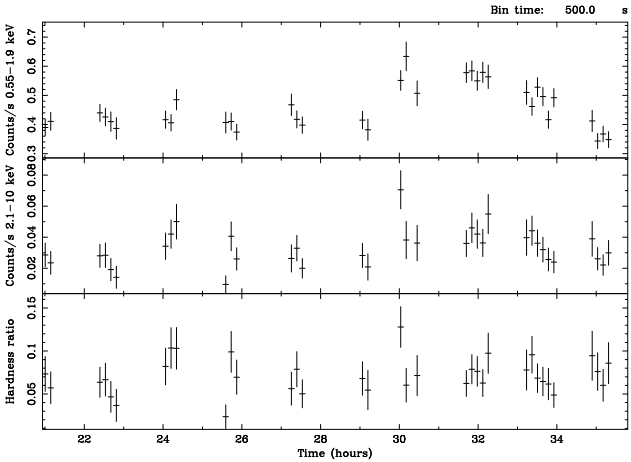}}
\caption{SIS0 light curves (not background subtracted) integrated in the 
energy 
bands 0.55-1.9\,keV (top), 2.1--10\,keV (middle), and the related hardness 
ratio 
(bottom). Observations were performed starting on 1993, May 7.}
\label{fig:ascalc}
\end{figure}

The light curves (Figure\,\ref{fig:ascalc}) were integrated considering two 
energy bands between 0.55--1.9\,keV and 2.1--10\,keV. The count rate is 
essentially constant for the first half of the observation, $\sim 0.4$\,cts 
for 
the SIS. The second half, on the contrary, shows an increase of $\sim 50$\% in 
particular in the softest band. However, the hardness ratio does not show 
significant variations due to the relatively poor statistics involved.

\subsection{ROSAT spectra}
\label{sec:rspec}

The high ROSAT count rate allowed us to perform a time--resolved spectral 
analysis. We applied the optically thin plasma codes of Raymond \& Smith 
(\cite{RS77}) and Mewe et al. (\cite{MKL96}, MEKAL). Both codes adopt 
optically--thin, collisional ionization equilibrium emissivity models. Since 
no 
significant differences were found in the results obtained with the two models 
we only report the MEKAL ones. The presence of absorbing material has been 
taken 
into account using the WABS component in XSPEC, which implements the Morrison 
\& 
McCammon (\cite{MM83}) model of X--ray absorption from interstellar material. 
Abundance variations in the source spectra have been modeled through a single 
parameter, the global ``metallicity'' $Z$, by assuming a fixed ratio between 
individual elemental abundances and the corresponding solar photospheric 
values 
as given by Anders \& Grevesse (\cite{AG89}). 

The errors on the counts have been taken into account by the Gehrels 
(\cite{Geh86}) approximation for data following the Poissonian statistics 
rather than the Gaussian statistics. Systematic errors were also considered 
for 
the ROSAT data as discussed later. The $\chi^2$ minimization statistics was 
applied through the paper. 

ROSAT spectra were extracted from the event files using the XSELECT/FTOOLS 
package and were rebinned to give at least 25 counts per bin. For all the 
observations a source spectrum was accumulated from a circular region of about 
$2 \div 3$\,arcmin, while a background spectrum was integrated in an annular 
region centered on the source with inner and outer radii of 4 and 11\,arcmin, 
respectively. We also use the spectra produced automatically by the WGA 
catalog 
analysis procedures. In both cases the derived fit parameters were essentially 
the same. Being \object{Gl\,355} a relatively bright and isolated X--ray 
source, 
the automatic extraction procedure used in the WGA catalog processing produced 
accurate results. 

The spectra for the individual ROSAT pointings were fitted with 1-- and 
2--temperature models, with variable $N_{\rm H}$ and metallicity $Z$. No 
attempt 
was made to vary the abundances of individual elements, since this is not 
warranted by the PSPC low--resolution.

For the observations with higher count rates (i.e. during the main part
of the flare) we did not get any satisfactory spectral fit taking into account 
the Poissonian errors only, and no significant improvements could be obtained 
by 
adding further thermal components. In low--resolution spectra, such as those 
provided by the PSPC, high $\chi^2_{\rm red}$ values are unlikely to be due to 
uncertainties in the model. Coronal spectra obtained by ROSAT have in fact 
been 
widely fitted successfully. We rather interpret this result as a consequence 
of 
the high statistics of our spectra during the flare evolution which leads to 
Poissonian errors comparable to or less than the other sources of uncertainty. 
In particular, a calibration error of $\sim 2\%$ is expected to affect on 
average each spectral channel and we took it into account (Fiore et al. 
\cite{FEMSW94}, Bocchino et al. \cite{BMS94}).

\subsection{ASCA spectra}
\label{sec:aspec}

The extracted ASCA spectra were rebinned to give at least 25 counts per 
bin. For the SIS detectors, we used the standard background data provided by 
the 
ASCA observatory team since the source almost completely fills the 1--CCD mode 
FoV. Standard screening was applied and spectra were accumulated in a region 
as wide as possible around the source: 3 and 2\,arcmin for SIS0 and SIS1, 
respectively. The GIS spectra were integrated in a region containing 98\% of 
the source flux. A background spectrum was accumulated from the outer regions 
of the FoV.

To avoid known ASCA calibration problems, the spectral analysis was restricted 
to the energy range 0.55-10\,keV for the SIS detectors and 0.6--10\,keV for 
the 
GIS detectors. In fact, spectral channels at higher energies are dominated 
by the noise due to the rapid effective area decrease, while SIS channels with 
energy less then 0.55\,keV have relevant calibration uncertainties (Dotani et 
al. \cite{DYR95}). No acceptable fit, in fact, was obtained including the 
low--energy channels both in terms of $\chi^2$ values and of the distribution 
of 
residuals.

We have performed spectral fits of ASCA data for the SIS0, SIS0+SIS1, GIS2, 
GIS2+GIS3, and SIS0+GIS2 datasets. SIS0 and GIS2 are the best calibrated 
detectors. The fits to these data were performed with 1--, 2-- or 
3--temperature 
models and free metallicity $Z$. Only in the case of the SIS spectra we have 
also performed the analysis with individual elemental abundances left free to 
vary. In all fits, in order to reduce the number of free parameters, the 
interstellar absorption $N_{\rm H}$ was fixed at $4 \times 
10^{19}$\,cm$^{-2}$, 
as suggested by the ROSAT analysis (Sect.\,\ref{sec:quie}). However, due to 
the 
harder ASCA energy band, the adopted value of the absorbing column does not 
affect the results.

\section{Results}
\label{sec:results}

\subsection{Quiescent Emission}
\label{sec:quie}

The ROSAT observations 200998 (the pre--flare observation), 201003, 201004, 
201005, the ASCA observation and the RASS data allow us to study the quiescent 
emission of \object{Gl\,355} on a long and a short time--scale. As already 
pointed out in Sect(s).\,\ref{sec:rpsps} and \ref{sec:aobs} on the basis of 
the 
light curves, both long-- and short--term variability is clearly present 
amounting to up a factor of $2 \div 3$ in flux. 

\begin{table*}[ht]
\begin{center}
\begin{tabular}{|lccccccc|}
\hline
Obs. & $N_{H}$ & $KT$ & EM & $Z$ & flux$_{0.1 - 2.4\,{\rm keV}}$  & 
$\chi_{\nu}^{2}$ 
& 
d.o.f. \\
   &($10^{19}$ cm$^{-2}$) &keV &$10^{52}$\,cm$^{-3}$ &$Z/Z_{\odot}$ &erg 
s$^{-1}$ cm$^{-2}$ & & \\
\hline
200998n00  &$4.20\pm^{3.16}_{2.63}$  &$0.63\pm^{0.08}_{0.08}$ &9.04  
&$0.08\pm^{0.04}_{0.03}$ &$1.43 \times 10^{-11}$ &0.22 &14 \\
201003n00  &$3.60\pm^{2.08}_{1.84}$  &$0.76\pm^{0.07}_{0.06}$ &14.53 
&$0.09\pm^{0.03}_{0.03}$ &$2.58 \times 10^{-11}$ &1.3 &14 \\
201004n00  &$4.70\pm^{1.88}_{1.72}$  &$0.88\pm^{0.11}_{0.09}$ &21.39 
&$0.05\pm^{0.03}_{0.02}$ &$3.5 \times 10^{-11}$ &1.2 &14 \\
201005n00  &$3.13\pm^{2.04}_{1.79}$  &$0.75\pm^{0.07}_{0.06}$ &14.93  
&$0.10\pm^{0.04}_{0.03}$ &$2.79 \times 10^{-11}$ &1.3 &14 \\
\hline
\end{tabular}
\end{center}
\caption{1T fits to the ROSAT spectra for the quiescent emission. 
Errors at 90\% confidence for the parameters of interest.}
\label{tab:rsmk1}
\end{table*}

We have tried to fit the ROSAT PSPC observations on Nov 1992 with the simplest 
possible model: a 1T model with free global metallicity. No observation could
be fitted with solar metallicity even assuming 2T and 3T models. As reported 
in 
Table\,\ref{tab:rsmk1}, the first model gives acceptable fits. The absorbing
column on the 
line of sight is not well constrained (errors of the order of 50\% of the best 
fit value, or larger), but the mean value is $\sim 4 \times 
10^{19}$\,cm$^{-2}$ 
and we could not obtain satisfactory fits with $N_{\rm H}$ lower than a few 
times $10^{19}$\,cm$^{-2}$. 

\begin{figure}
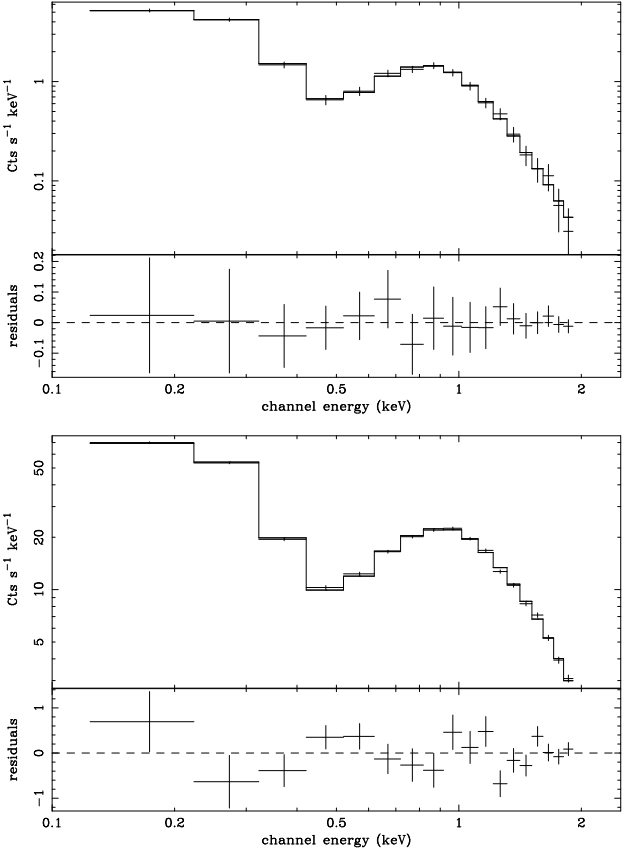

\begin{tabular}{c}
\rotatebox{270}{\resizebox{5.5cm}{!}{\includegraphics{ros_spec.ps}}} \\
\rotatebox{270}{\resizebox{5.5cm}{!}{\includegraphics{ros_spec2.ps}}} \\
\end{tabular}
\caption{(Top) ROSAT spectrum for observation 200998 (the pre--flare 
observation). The best fit 1T model (Table\,\ref{tab:rsmk1}) and fit residuals 
are also shown. (Bottom) ROSAT spectrum for observation 200999 (the flare 
peak). 
The best fit 2T model (Table\,\ref{tab:rsmk2}) and fit residuals are also 
shown.}
\label{fig:rosspec}
\end{figure}

The metal abundance is highly subsolar ($Z/Z_\odot = 0.03 \div 0.14$ at the 
90\% 
confidence level) and is confirmed by the analysis of the ASCA spectra 
discussed 
below.  The best--fit temperature is $\sim 0.7$\,keV with a moderate 
hardening for observation 201004 coincident with the small flare apparently 
superimposed to the tail of the main event (Fig.\,\ref{fig:totlc}). The flux 
in 
the 0.1--2.4\,keV energy band ranges from $\sim 1.4$ to $\sim 3.5 \times 
10^{-11}$\,erg\,cm$^{-2}$s$^{-1}$, directly linked to the EM variations from 
$\sim 9$ to $\sim 21 \times 10^{52}$\,cm$^{-3}$. All ROSAT observations 
outside 
of the flare can thus be fitted by 1T models (see the best fit for 
observation 200998 in Fig.\,\ref{fig:rosspec}, top panel). The spectral 
parameters derived for the pre-- and post--flare emissions are comparable and 
the moderate (a factor of $2 \div 3$) flux variability seems to be due mainly 
to 
EM variations which in turn might be due to changes in either the volume or 
the 
density of the emitting regions.

\begin{table*}
\begin{center}
\begin{tabular}{|ccccccccc|}
\hline
Camera    & KT$_1$  & EM$_1$  & KT$_2$  & EM$_2$ & Z & flux$_{0.55-10\,{\rm 
keV}}$    
& $\chi_{\nu}^2$ & d.o.f. \\
          & keV     & $10^{52}$\,cm$^{-3}$ & keV & 
$10^{52}$\,cm$^{-3}$ & $Z/Z_\odot$ & erg\,s$^{-1}$\,cm$^{-2}$ & & \\
\hline
SIS0  & $0.65\pm^{0.05}_{0.09}$ &  4.44 & $1.30\pm^{0.34}_{0.27}$  & 3.23 & 
$0.15\pm^{0.07}_{0.05}$ & $1.10\times10^{-11}$ & 0.9 & 103 \\
SIS0+SIS1 & $0.62\pm^{0.01}_{0.07}$ & 4.44 & $1.17\pm^{0.21}_{0.15}$ & 3.63 & 
$0.13\pm^{0.05}_{0.03}$ & $1.10\times10^{-11}$ & 0.9 & 203 \\
GIS2  & $0.87\pm^{0.10}_{0.09}$ & 6.46 &   -  &  - & $0.11\pm^{0.07}_{0.04}$  
& $0.81\times10^{-11}$ & 0.8 & 111 \\
GIS2+GIS3 & $0.89\pm^{0.07}_{0.06}$ & 6.46 &  - & -  & 
$0.12\pm^{0.05}_{0.03}$ & $0.83\times10^{-11}$ & 0.8 & 245 \\
SIS0+GIS2 & $0.65\pm^{0.04}_{0.06}$ & 4.44 & $1.39\pm^{1.60}_{0.27}$ & 2.42 & 
$0.17\pm^{0.08}_{0.05}$ & $1.00\times10^{-11}$  & 1.1 & 217 \\  
\hline
\end{tabular}
\end{center}
\caption{1T and 2T fits to ASCA spectra. For simultaneous fits the 
normalizations are kept the same for both SIS and GIS detectors. 
Errors at 90\% confidence for the parameters of interest.}
\label{tab:rsmk2zv_asca}
\end{table*}

Considering the ASCA observation on May 1993 1T spectral fits gave 
satisfactory 
results only for the low resolution GIS detectors. The best fit 
temperature appears slightly harder (but comparable within the errors) than 
that 
one derived from the ROSAT analysis, while the metal abundance is essentially 
the same (Table\,\ref{tab:rsmk2zv_asca}). 

\begin{table*}
\begin{center}
\begin{tabular}{|lccccccccc|}
\hline
Obs. & $N_{H}$ & $KT_{1}$ & EM$_1$ & $KT_{2} $ & EM$_2$ & $Z$ & flux$_{0.1 - 
2.4 
{\rm keV}}$  & $\chi_{\nu}^{2}$ & d.o.f. \\
     & $(10^{19}$ cm$^{-2}$) &keV &$10^{52}$\,cm$^{-3}$ & keV & 
$10^{52}$\,cm$^{-3}$ & $Z/Z_{\odot}$ &erg s$^{-1}$ 
cm$^{-2}$ & & \\
\hline
200999n00  &$1.87\pm^{1.76}_{1.54}$  &$0.71\pm^{0.23}_{0.26}$ &8.76  
&$4.44\pm^{n.c.}_{1.53}$ &114.30 &$0.55\pm^{0.80}_{0.45}$ &$3.11 \times 
10^{-10}$ &1.0 &12 \\
201000n00  &$2.80\pm^{2.31}_{2.20}$  &$0.53\pm^{0.20}_{0.19}$ &4.84  
&$2.69\pm^{2.29}_{0.76}$ &41.17 &$0.62\pm^{1.20}_{0.38}$ &$1.28 \times 
10^{-10}$ 
&0.9 &12 \\
201001n00  &$2.83\pm^{2.21}_{2.07}$  &$0.61\pm^{0.23}_{0.29}$ &8.88  
&$1.65\pm^{n.c.}_{0.40}$ &25.51 &$0.28\pm^{0.40}_{0.17}$ &$8.45 \times 
10^{-11}$ 
&0.9 &12 \\
201002n00  &$2.72\pm^{2.71}_{2.32}$  &$0.61\pm^{0.17}_{0.29}$ &7.67  
&$1.75\pm^{n.c.}_{1.43}$  &10.90  &$0.21\pm^{0.35}_{0.12}$ &$4.22 \times 
10^{-11}$ &0.5 &12 \\
\hline
\end{tabular}
\end{center} 
\caption{2T ROSAT MK fits for the flare emission. The metal abundance is free 
to 
vary in solar proportion. Errors at 90\% confidence for the parameters of 
interest.}
\label{tab:rsmk2}
\end{table*}

Acceptable fits to the SIS spectra were obtained only with the addition of 
a second thermal component, with the global metallicity free to vary 
(Table\,\ref{tab:rsmk2zv_asca}). The metal abundance is comparable to that 
obtained from the analysis of the ROSAT data and is well below solar. 
Since ASCA is much more effective than ROSAT in measuring metal abundances, 
this result clearly shows that a low metal abundance is indeed needed to model 
the corona of \object{Gl\,355} in spite of the fact that this is a very young 
star with presumably solar photospheric abundances. As expected, the 
introduction 
of a second component gives a lower value for the cooler temperature, while 
the 
hotter temperature is around $1.3$\,keV and its EM is about 25--50\% 
lower than that one of the cooler component. The higher temperature 
derived for the ASCA 2T fit is due to the much larger energy range involved, 
which makes ASCA sensitive to hotter plasma than the ROSAT PSPC. 

\begin{table}
\begin{center}
\begin{tabular}{|cc|cc|cc|}
\hline
 & & SIS0 & SIS0+SIS1 \\
\hline        
$KT_1$ & (keV) & $0.64\pm^{0.00}_{0.03}$ & $0.63\pm^{0.02}_{0.030}$ \\ 
EM$_1$ & ($10^{52}$\,cm$^{-3}$) & 3.96 & 4.44 \\ 
$KT_2$ & (keV) & $2.10\pm^{0.70}_{0.30}$ & $2.00\pm^{0.40}_{0.30}$ \\ 
EM$_2$ & ($10^{52}$\,cm$^{-3}$) & 1.53 & 1.61 \\ 
O$_{(FIP=13.6\,eV)}$ &  & $0.47\pm^{0.30}_{0.19}$ & $0.39\pm^{0.17}_{0.13}$ \\ 
Ne$_{(FIP=21.6\,eV)}$ &  & $0.65\pm^{0.36}_{0.24}$ & $0.55\pm^{0.21}_{0.16}$ 
\\ 
Mg$_{(FIP=7.6\,eV)}$  &  & $0.32\pm^{0.23}_{0.15}$ & $0.24\pm^{0.13}_{0.10}$ 
\\ 
Si$_{(FIP=8.2\,eV)}$  &  & $0.33\pm^{0.18}_{0.15}$ & $0.31\pm^{0.12}_{0.10}$ 
\\ 
S$_{(FIP=10.4\,eV)}$ &  & $0.50\pm^{0.42}_{0.38}$ & $0.41\pm^{0.27}_{0.25}$ \\
Fe$_{(FIP=7.9\,eV)}$  &  & $0.17\pm^{0.09}_{0.05}$ & $0.15\pm^{0.05}_{0.04}$ 
\\ 
$\chi^2_{\nu}$ & (d.o.f. = 98) & 0.7 & 0.7 \\ 
flux$_{0.55-10\,{\rm keV}}$ & erg\,s$^{-1}$\,cm$^{-2}$ & $1.2\times10^{-11}$ & 
$1.2\times10^{-11}$ \\
\hline
\end{tabular}
\end{center}
\caption{2T fits to ASCA spectra with individual abundances free to
vary. For simultaneous fits the normalizations are 
kept the same for both SIS cameras. 
Errors at 90\% confidence for only one parameter of interest.}
\label{tab:rsmk2zff_asca}
\end{table}

In order to study in detail the chemical composition of stellar coronal 
plasmas 
several authors have fitted ASCA spectra with thermal models in which the 
elemental abundances are allowed to vary individually, rather than in a fixed 
ratio with respect to the solar values (e.g. White et al. \cite{WAD94}, Drake 
et 
al. \cite{DSWS94}, Mewe et al. \cite{MKWP96}, Tagliaferri et al. \cite{TCF97}, 
Ortolani et al. \cite{OMP97}). We have tried the same approach for the SIS0 
and 
SIS0+SIS1 datasets (Table\,\ref{tab:rsmk2zff_asca}) considering as free 
parameters only the ions that contribute most to line emission in the ASCA 
passband. These include O, Ne, Mg, Si, S and Fe, whose abundances can be 
sufficiently well constrained. Similar fits made by adding N, Ar and Ca as 
free 
parameters resulted in essentially unconstrained abundances for these 
elements. 
The abundances of all other elements were frozen to their solar values (see 
Mewe 
et al. \cite{MKvVT97} for a discussion).

Satisfactory fits were obtained with 2T models 
(Table\,\ref{tab:rsmk2zff_asca}). 
The hotter temperature is harder than that obtained with 2T fits and a 
variable 
global metallicity, while the ratio between the two EM is lower by $\sim 
40$\%. 
This is likely due to the redistribution of the best fit element abundances 
which, although all below solar, show a significant lower iron abundance with 
respect to solar than the other elements. Note however that the fit with 
variable individual abundances is not statistically better than the one with a 
single global metallicity.

Finally, we have tried to fit the SIS spectra with 3--temperature (3T) models 
and solar abundances. These fits still did not give satisfactory results. The 
addition of a third component simply redistributed the temperature over a 
wider 
range. If the global metallicity is left free to vary both the SIS0 and the 
SIS0+SIS1 data sets gave satisfactory results but the introduction of a 
further 
thermal component is not formally required by the fit.

\subsection{Flare analysis}
\label{sec:flare}

In order to analyze the flare emission it is necessary to separate it from the 
quiescent emission of the corona. In the previous section we have shown 
that the definition of a ``quiescent spectrum'' for \object{Gl\,355} is not 
straightforward since a significant amount of variability is present. 
Considering the ROSAT observations 200999, 201000, 201001 and 201002, i.e. the 
observations performed during the flare, no satisfactory fit could be obtained 
with 1T models. Even subtracting the ROSAT observation 200998, or an average 
of 
the three post--flare observations 201003, 201004 and 201005, no adequate fit 
could be obtained. We thus performed 2T fits with the global metallicity 
either 
fixed to the solar value or left free to vary. With solar metallicity the fits 
are worse than the corresponding fits with 1T and global metallicity free to 
vary. However, if we allow the metal abundance to vary, the 2T fits give 
acceptable results (Table\,\ref{tab:rsmk2} and Fig.\,\ref{fig:rosspec}, bottom 
panel), but the hotter component is badly constrained. No better constraining 
can be obtained by freezing the absorbing column and/or fixing the metal 
abundance to the quiescent value. The intense dynamic evolution of the flare 
prevents a detector with a limited energy band as the ROSAT PSPC to constrain 
the hot component. 

\begin{figure}
\resizebox{\hsize}{!}{\includegraphics{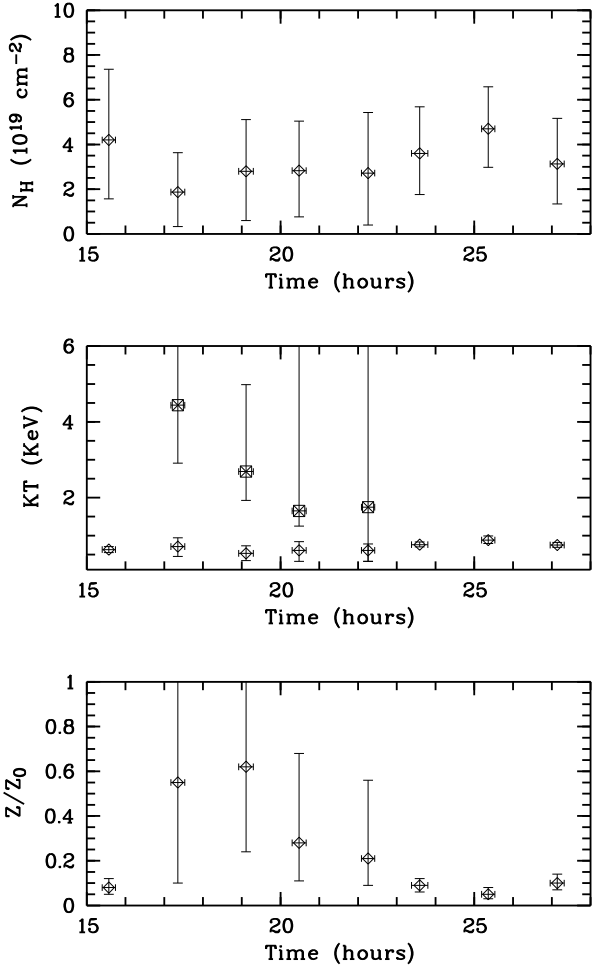}}
\caption{1 and 2--T best fit parameters for the ROSAT PSPC observations. The 
first observation and the last three were fitted by 1T models, while the 
observations performed during the flare were fitted with 2T models 
(see Fig.\,\ref{fig:totlc}).}
\label{fig:T}
\end{figure}

As shown in Fig.\,\ref{fig:T}, the temperature of the flare cooler component 
is 
essentially constant around 0.6--0.7\,keV, in full agreement with the 
temperature derived for the quiescent emission. 

The absorbing column density is also essentially constant during the flare, 
i.e. with no increase at the flare onset or at the peak which, if present,  
could be interpreted as evidence for a mass ejection. The metal abundance 
close 
to the flare peak is not well constrained but a hint (admittedly very weak) 
for 
an increase during the flare evolution seems to be present 
(Fig.\,\ref{fig:T}). 

In any case a hot component is clearly needed, although being not well 
constrained due to the limited ROSAT energy range. The flare event is totally 
due to hot plasma superimposed to the quiescent corona. 

We note in passing that the observation 201000 shows for the low temperature 
component the lowest EM among the flare observations, about half of the values 
obtained for observations 200999, 201001 and 201002. The temperature is 
slightly 
reduced. These two factors likely explain why the hardness ratio for this 
observation is the hardest during the flare (cf. Table\,\ref{tab:rosatobspar} 
and Fig.\,\ref{fig:LC}). This phenomenum may also be related to the presence 
of 
heating during the decay phase of the flare discussed in 
Sect.\,\ref{sec:model}.

\section{Flare modeling}
\label{sec:model}

\subsection{Loop modeling}
\label{sec:modela}

The flare observed by ROSAT has been analyzed considering the so called 
hydrodynamic decay--sustained heating scenario (Reale et al. \cite{RBPSM97}), 
which assumes that the flaring plasma is confined in a closed loop structure 
whose geometry does not change during the event. This method simultaneously 
yields estimates for the size of the flaring loop as well as for the presence 
and time scale of heating during the decay phase of the flare. The method uses 
the slope of the locus of points in the temperature vs. density diagram of the 
flare decay phase which, from hydrodynamic simulations of X--ray flares, has 
been found to be a good diagnostics for the presence of sustained heating. 
Under 
the assumption that the loop volume remains constant during the flare, the 
square root of the emission measure can be used as an indicator for the 
density. 
We also assume that the flare loop length is small in comparison with the 
coronal pressure scale height i.e. that isobaric conditions hold for the 
plasma 
in the flaring loop (we will verify the correctness of this assumption a
posteriori). For the present flare, we make the further assumption that the 
hot 
component found in the 2T fits discussed in the previous section is indeed 
responsible for the flare event (i.e. that the cool component contributes only 
to the quiescent emission). 

Detailed hydrodynamical simulations (Peres et al. \cite{PSVR82}, Betta et al. 
\cite{BPRS97}) show that flares decay approximately along a straight line in 
the $\log \sqrt{EM} - \log T$ diagram, and that the value of the slope $\zeta$ 
of the decay path is related to the ratio between the observed decay time 
$\tau_{\rm lc}$ of the light curve and the thermodynamic cooling time of the 
loop $\tau_{\rm th}$ in the absence of heating during the flare decay. This 
allows deriving the length of the flaring loop as a function of observable 
quantities. Since the characteristics of the observed decay depend on the 
instrument response, the specific relationships to be used depend on the 
instrument and must be appropriately calibrated. An application of this 
technique to stellar flares observed with the ROSAT PSPC, and the appropriate 
calibrations, are given by Reale \& Micela (\cite{RM98}) and Favata et al. 
(\cite{FMRSS00}). See also Pallavicini et al. (1999) for a critical discussion
of the method.

The theoretical thermodynamic decay time $\tau_{\rm th}$ (sec) of a closed 
coronal loop with semi--length $L$ (cm), and maximum temperature $T_{\rm max}$ 
(K) in the absence of sustained heating is (Serio et al. \cite{SRJSS91}):
\begin{equation}
\tau_{\rm th} = \frac{\alpha L}{\sqrt{T_{\rm max}}}
\end{equation}
where $\alpha = 3.7 \times 10^{-4}$\,cm$^{-1}$\,s\,K$^{1/2}$. By means of
a grid of hydrostatic loop models, an empirical relationship linking the 
loop maximum temperature $T_{\rm max}$ to the temperature $T_{\rm obs}$ (K) 
has been derived:
\begin{equation}
T_{\rm max} = 0.13 \times T_{\rm obs}^{1.163}.
\label{eq:tmax}
\end{equation}

In the present case, we cannot be sure that the flare maximum was 
observed but we can assume that the observation 200999 is not too far from the 
real flare peak. The observed peak temperature is $51.5 \pm ^{n.c.}_{17.7} 
\times 10^6$\,K and therefore the actual maximum temperature from 
Eq.\,(\ref{eq:tmax}) is $\sim 121 \times 10^6$\,K. 

\begin{figure}
\resizebox{\hsize}{!}{\includegraphics{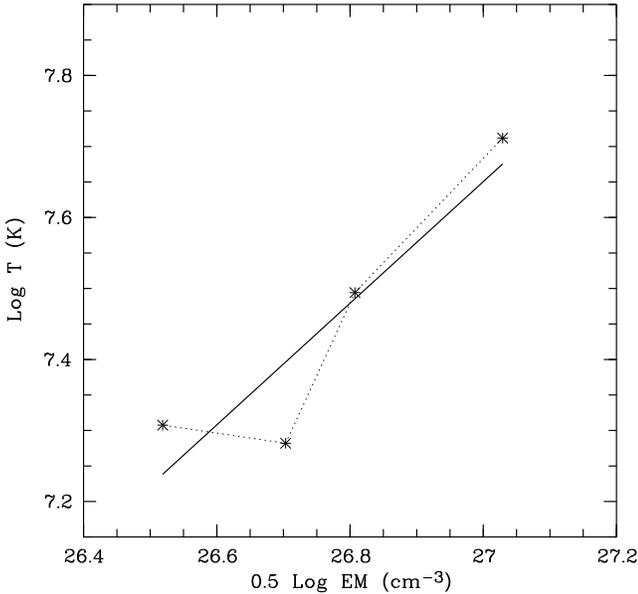}}
\caption{The evolution of the flare decay in the $\log T$ vs. $\log \sqrt{EM}$ 
plane. The dotted line connects points in temporal order while the solid line 
is 
the best fit to the decay.}
\label{fig:EMT}
\end{figure}

The ratio between $\tau_{lc}$ and $\tau_{th}$ is a function of the slope 
$\zeta$ 
in the $\log \sqrt{EM} - \log T$ (see Fig.\,\ref{fig:EMT}). The best fit for 
$\zeta$ with the present data is $0.86 \pm 0.27$. 

Applying the relation reported for the ROSAT data by Favata et al.
(\cite{FMRSS00}) valid for $0.4 \le \zeta \le 1.7$:
\begin{equation}
\frac{\tau_{\rm lc}}{\tau_{th}} = F(\zeta) = \frac{3.67}{\zeta/0.3 - 1.0}+1.61
\label{eq:fz}
\end{equation}
we get $F(\zeta) = 3.6 \pm 0.9$ with the value of the slope $\zeta$ derived 
above. Very low values of $\zeta$ mean that the flare decay is entirely 
driven by the sustained heating, so that the thermodynamic cooling time, 
$\tau_{\rm th}$, cannot be determined in a reliable way, and $L$ does not 
depend any more on $\tau_{\rm lc}$. On the other extreme, no sustained heating 
occurs and $\tau_{\rm lc} \sim \tau_{\rm th}$. The case discussed here is in 
the 
right regime where to apply Eq.\,(\ref{eq:fz}). Our result shows that a large 
amount of sustained heating is actually present in line with the results 
obtained for other flares (Reale \& Micela \cite{RM98}; Ortolani et al. 
\cite{OPMRW98}, Favata \& Schmitt \cite{FS99}; Favata et al. \cite{FRMSM00}; 
Favata et al. \cite{FMR00}; Favata et al. \cite{FMRSS00}; Maggio et al. 
\cite{MPRT00}, Franciosini et al. \cite{Fetal01}). 

The expression for the loop semi--length $L$ is:
\begin{equation}
L = \frac{\tau_{\rm lc}\sqrt{T_{\rm max}}}{\alpha F(\zeta)},
\label{eq:loop}
\end{equation}
For the present event, the decay e--folding time, $\tau_{\rm lc}$ is $10.1 
\pm 0.5$\,ks, and with the values above for $T_{max}$ and $\zeta$, the loop 
semi--length turns out to be $L = 83 (\pm 22) \times 10^{9}$\,cm or, in 
units of the stellar radius, $L \sim 1.5\,R_*$. The uncertainty on $L$ is 
given 
by the sum of the propagation of the errors on the observed parameters 
$\tau_{\rm lc}$ and $\zeta$ with the standard deviation of the difference 
between the true and the derived loop lengths. The latter uncertainty, 
estimated 
by applying the method to spatially resolved observations of solar flares, 
amounts to $\approx 20\%$ and usually dominates on the other error sources. 
The loop length derived here could be somewhat underestimated if the flare 
peak 
occurred earlier than observation 200999, i.e. at a time when we had a data 
gap.
The above relationships have been derived in the assumption of negligible 
interstellar absorption ($N_{\rm H} \sim 0$) and solar metallicity, but the 
corrections are small provided $N_{\rm H} \le 10^{20}$\,cm$^{-2}$ (Reale \& 
Micela \cite{RM98}) and the metal abundance is at least of the order 
$Z/Z_\odot 
\sim 0.1$: these conditions are satisfied in our case. 

The derived loop size is small in comparison with the pressure scale height, 
thus justifying the initial assumption that the flaring loops are isobaric.
In fact, the pressure scale height is $H = 2KT/\mu g$, where $T$ is the
plasma temperature in the loop, $\mu$ is the molecular weight and $g$ is
the surface gravity of the star. With $\log g/g_\odot \sim 0.07$
and $\mu \sim 0.6$ we get $H \sim 6 \times 10^{11}$\,cm. 

The average plasma density can be estimated as:
\begin{equation}
n_e \sim \sqrt{\frac{EM}{V}},
\label{eq:em}
\end{equation}
where $EM$ is the emission measure derived from the fit,  
and $V$ the loop volume. 
Assuming $\beta = 0.1 \div 0.3$ for the ratio between the radius of the loop 
cross--section and its semi--length (typical values for solar coronal loops, 
Golub et al. \cite{GMR80}) we obtain $V = 2 \pi L^3 \beta^2 = 3.6 \div 33 
\times 
10^{31}$\,cm$^3$ (considering the uncertainties on $\beta$ and $L$). 
Eq.\,(\ref{eq:em}) thus yields 
$n_{\rm e} = 2 \div 0.7 \times 10^{11}$\,cm$^{-3}$.
The corresponding pressure 
$p_e = 2n_{\rm e} k T_{\rm max} = 6.7 \div 2.3 \times 10^3$\,dyne cm$^{-2}$.

Using the scaling laws of Rosner et al. (\cite{RTV78}) applicable to static 
loops $T_{\rm max} = 1.4 \times 10^3 (p_0L)^{1/3}$\,K, where $p_0$ is the
pressure at the base of the loop, we derive, at the temperature peak, $p_0 = 
7.7\times 10^3$\,dyne cm$^{-2}$. The similarity of $p_0$ and $p_e$, for the 
adopted range of values for $\beta$, implies that the plasma at the flare peak 
is not far from quasi-static conditions, i.e. the loop is almost completely 
filled with plasma evaporated from the chromosphere, as a consequence of the 
heating. 

\begin{table}
\begin{tabular}{|lccc|}
\hline
Obs.               & $L_{X_{(0.1 \div 2.4\,{\rm KeV})}}$ &  $L_{X_{(0.1 \div 
2.4\,{\rm KeV})}}$ & \\
                   &  erg\,s$^{-1}$   &    erg\,s$^{-1}$   &  $L_{\rm 
flare}/L_X$ \\
                   & total            & flare only &  \\
\hline
RASS               & $8.50 \times 10^{29}$    &   & \\
200998n00          & $5.77 \times 10^{29}$    &   & \\
200999n00          & $1.25 \times 10^{31}$    & $1.09 \times 10^{31}$ & 0.87 \\
201000n00          & $5.10 \times 10^{30}$    & $4.16 \times 10^{30}$ & 0.82 \\
201001n00          & $3.40 \times 10^{30}$    & $2.34 \times 10^{30}$ & 0.69 \\
201002n00          & $1.70 \times 10^{30}$    & $0.93 \times 10^{30}$ & 0.55 \\
201003n00		& $1.00 \times 10^{30}$    &   & \\
201004n00		& $1.35 \times 10^{30}$    &   & \\
201005n00		& $1.08 \times 10^{30}$    &   & \\
\hline
\end{tabular} 
\caption{X--ray flare luminosity evolution. For comparison the luminosity 
detected during the RASS observation and for the pre-- and post--flare 
quiescent 
emissions are also reported. In Column\,2 the total luminosity from 
\object{Gl\,355} is reported; in Column\,3 the luminosity of the flare 
component 
only and, finally, in Column\,4, the fraction of the total luminosity due to 
the 
flare component.}
\label{tab:lum}
\end{table}

It is interesting to compute the coronal magnetic field strength required to 
keep confined a plasma with the density reported here. Equating the magnetic 
pressure to the gas pressure in the flaring loop, $B^2/8 \pi \sim p_e$, we get 
a 
value of $\sim$ 300\,G in the corona. If we assume that the magnetic field has 
a 
dipolar structure, i.e. it scales with height with a $r^{-3}$ law, the 
expected 
magnetic field at the stellar surface in the flaring region comes out to be of 
the order of 2500\,G. Recently, Donati (\cite{D99}) studied the magnetic 
cycles 
of \object{HR\,1099} and \object{Gl\,355} by means of Zeeman--Doppler imaging, 
and reported evidences for large--scale magnetic fields on \object{Gl\,355} 
with strengths of up to a few hundred Gauss. This discrepancy could mean 
that either the large scale magnetic field studied by Donati (\cite{D99}) 
is not related to the localized fields responsible for large flares or that 
the 
structure of these localized fields is not dipolar. In any case, conventional 
Doppler imaging measurements show that magnetic fields of the order of a few 
thousands Gauss are not uncommon on late--type active stars as 
\object{AD\,Leo} 
(see for instance Saar \& Linsky \cite{SL85} and Linsky \& Saar 
\cite{LS87}). 
We stress, however, that within the hypothesis of dipolar fields, the inferred 
strength of the photospheric magnetic field is not significantly affected by 
the 
derived flare loop size because a smaller loop implies a higher gas and 
magnetic 
pressure in the corona, but also a smaller difference between the strength of 
the photospheric and coronal magnetic fields. For instance, if the flare loop 
were a factor ten smaller, we would still require a photospheric magnetic 
field 
of about 2200\,G in the dipole approximation.

Finally, we have computed, for each time interval during the flare, the 
X--ray luminosity in the $0.1 \div 2.4$\,keV band (see Table\,\ref{tab:lum}). 
A detailed determination of the energy balance of the present flare is not 
possible given the lack of multi-wavelength coverage and velocity information 
which could help assessing the plasma kinetic energy. The \object{Gl\,355} 
bolometric luminosity is $\sim 1.8 \times 
10^{33}$\,erg\,s$^{-1}$. The total energy radiated by the flare in X--rays in 
the band $0.1 \div 2.4$\,keV is $E \sim 9 \times 10^{34}$ over $\sim 19$\,ks, 
and is equivalent to $\sim 0.4$\% of the star's bolometric energy output 
during 
the same interval. The observed peak 
flare luminosity was $\sim 1$\% of the star's bolometric luminosity.

\subsection{``Order of magnitude" estimates}

It is interesting to compare the above results, based on self-consistent
hydrodynamic loop models, with simple order of magnitude estimates of the 
radiative and conductive cooling times for a single flaring loop with no 
additional heating in the decay phase. This approach is the one that has been 
most commonly used in previous analyses of stellar flares. We follow here the 
formalism of Pallavicini (\cite{P95b}; see also Pallavicini et al. 
\cite{PTS90}). For a loop of density $n_e$ and temperature $T$, the radiative 
cooling time is given by:
\begin{equation}
\tau_{\rm rad} = \frac{3kT}{n_eP(T)}
\end{equation}
where $k$ is the Boltzmann constant and P(T) is the radiative loss function 
for unit emission measure which can be approximated as $P(T) \simeq 
10^{-24.73}T^{0.25}$\,erg\,cm$^{3}$\,s$^{-1}$ for temperatures higher than 
20\,MK (Mewe et al. \cite{MGv85}). If the flare cools predominantly by 
radiation, the above equation, together with the flare coronal temperature 
derived from the spectral fits (see Table\,6), allows deriving a value for the 
flare density (or only an {\it upper} value to it, if conduction is not 
negligible). With $T= 51.5 \times 10^6$\,K, this gives $n_e = 1.3 \times 
10^{11}$\,cm$^{-3}$, which, from Eq.\,(\ref{eq:em}), results in a volume (or 
in 
a {\it lower} limit to it) of $6.8 \times 10^{31}$ cm$^3$. With $\beta$ in the 
range $0.1 \div 0.3$, the loop semi--length is thus $L \sim 100 \div 50 \times 
10^9$\,cm. i.e. slightly larger than the stellar radius and comparable to the 
values inferred above from the hydrodynamic modeling. 

As discussed by Haisch (1983, see also Pallavicini 1995), if conduction losses 
are not negligible, but are comparable to radiative losses, it is possible to 
determine uniquely the loop length (as opposite to determining only an {\it 
upper} limit to it), under the assumption again that there is no heating 
during 
the flare decay. In fact, the conductive cooling time is given by:
\begin{equation}
\tau_{\rm cond} = 4.8 \times 10^{-10} \frac{n_e L^2}{T^{2.5}}
\end{equation}
which, for $\tau_{\rm cond} \sim \tau_{\rm rad}$, gives $L = 56 \times 10^{9}$ 
cm, which is again comparable to the stellar radius in the case of 
\object{Gl\,355}. Finally, we note that the so-called quasi-steady cooling 
model 
(van den Oord and Mewe \cite{vM89}), often used in modeling stellar flares, 
coincides, for the radiative loss function $P(T)$ assumed above, to a ratio 
$\tau_{\rm rad} / \tau_{\rm cond} = 0.18$, i.e. to the case of radiation 
dominated cooling, although conduction is not completely negligible. Hence, 
the 
loop length estimated from the quasi-static cooling model cannot be too 
different from that estimated from radiative cooling alone (Pallavicini 
\cite{P95b}).

Due to the similarity between the estimates of the \object{Gl\,355} flare loop 
size performed with the Reale et al. (\cite{RBPSM97}) approach and with simple 
``order of magnitude" considerations, we also analyzed four flares from 
\object{Algol} (Favata et al. \cite{FMRSS00}), three flares from 
\object{AB\,Dor} (Franciosini et al. \cite{Fetal01}), one more flare from 
\object{AB\,Dor} (Ortolani et al. \cite{OPMRW98}), six flares from 
\object{AD\,Leo} (Favata et al. \cite{FMR00}) and one flare from 
\object{EV\,Lac} (Favata et al. \cite{FRMSM00}).

\begin{figure}
\resizebox{\hsize}{!}{\includegraphics{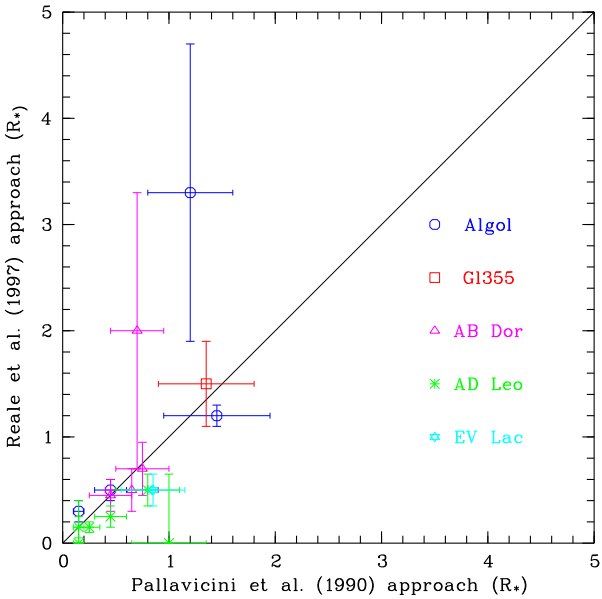}}
\caption{Comparison between flares loop sizes determined by the methodology 
developed by Reale et al. (\cite{RBPSM97}) and by the ``order of magnitude" 
estimates by Pallavicini et al. (\cite{PTS90}).}
\label{fig:flares}
\end{figure}

The result of this comparison is shown in Fig.\,\ref{fig:flares}. In spite of 
the fact that simple orders of magnitude estimates are clearly based on 
physically wrong assumptions (since evidences of sustained heating during the 
decay of the considered flares have been usually reported), they 
provide values for the flare volume and density which are not too dissimilar 
from those derived from the hydrodynamic decay-sustained heating model of 
Reale 
et al. (\cite{RBPSM97}). Moreover, in many cases, the loop semi--length derived
with both methods is large and comparable to the stellar radius, a situation 
quite different from that encountered in the case of solar flares. The large 
flare size is clearly needed to provide the large energy release of stellar 
flares with respect to solar ones, without requiring unrealistically high 
values 
of the magnetic field.

Several factors are probably conspiring here to give us this surprising 
coincidence. On one hand it is well known that if there is evidence 
of heating in the flare decay, the use of the observed quantity
$\tau_{\rm lc}$ instead of the theoretical $\tau_{\rm th}$ to estimate
the flare sizes, intrinsically produces too large loop lengths.
On the other hand, if we assume that there is no heating, the density
estimated from radiative cooling only should be higher than the true
values since the flare also cools by conduction and the volume derived
is correspondingly lower.  

Within the limits of the small sample analyzed here, no trend seems to emerge 
for flare size with respect to the star parameters as spectral types, rotation 
rates, etc.

\section{Summary and conclusions}
\label{sec:concl}

The spectral analysis of the ROSAT and ASCA data show that this young star has 
coronal metal abundances strongly sub--solar. Although there are no direct 
measurements of the photospheric abundances of \object{Gl\,355}, it is 
expected 
to be solar. Thus, as in the case of other young objects, e.g. 
\object{AB\,Dor} 
(Mewe et al. \cite{MKWP96}) and \object{HD\,35850} (Tagliaferri et al. 
\cite{TCF97}), we have a star that shows coronal abundances much lower than 
the 
photospheric ones. This is also true for other more evolved stars, as e.g. 
\object{II\,Peg} (Mewe et al. \cite{MKvVT97}; Covino et al. \cite{CTPMP00}). 
These ROSAT, ASCA and {\it Beppo}SAX findings are now confirmed with the 
gratings observations of Chandra for \object{HR\,1099} (Drake et al. 
\cite{DBK01}) and of XMM-Newton for \object{AB\,Dor}, \object{Castor} and 
\object{HR\,1099} (Brinkmann et al. \cite{BBG01}, G\"udel et al. \cite{GAB01}, 
\cite{GAM01}).

Besides the strong flare detected by ROSAT, variability of a factor of 2-3 has 
been detected in both ROSAT and ASCA light curves of \object{Gl\,355}. This 
variability is mainly due to EM changes, while the temperature of the 
quiescent 
corona remains approximately constant. For the ROSAT data, outside the flare, 
the coronal plasma is well represented by a single temperature model with very 
low metal abundances. For the ASCA data, due to the harder energy band, a 
second 
harder component is required. 

Applying the relation $N_{\rm H} \sim 0.07$\,cm$^{-3}$ from Paresce 
(\cite{P84}), and assuming a distance of $\sim 18$\,pc for \object{Gl\,355}, 
we 
obtain $N_{\rm H} \sim 4\times10^{18}$\,cm$^{-2}$. However, acceptable fits to 
the ROSAT data could only be obtained with $N_{\rm H}$ an order of magnitude 
higher. Fruscione et al. (\cite{FHJW94}), compiling a list of measured 
hydrogen 
column densities for stars mainly in the solar neighborhood, showed that for 
an 
object at slightly less than 20\,pc $N_{\rm H}$ of the order of $10^{19}$ are 
possible (their Fig.\,3). It has also been suggested (Rodon\'o et al. 
\cite{RPL99}) that extra--absorption might occur in some stars owing to the 
presence of neutral hydrogen in the circumstellar environment. 

The most interesting feature in the \object{Gl\,355} ROSAT data is of course 
the 
strong flare detected. The coronal spectrum during the flare can be 
represented 
with a 2-T model, with the cooler component compatible with the quiescent one. 
From these fits we have a weak indication of an increase of the metal 
abundance 
during the flare although the large error bars do not allow for a strong 
claim. 
A change of the metal abundance value during strong flare was detected for 
various sources and with different satellites (e.g. \object{Algol} with ROSAT 
and {\it Beppo}SAX: Ottmann \& Schmitt \cite{OS96}, Favata \& Schmitt 
\cite{FS99}; \object{AB\,Dor} with XMM-Netwon: G\"udel et al. \cite{GAB01}). A 
separate issue is whether a variation of the column absorption could occur 
during the flare. Enhancements of the hydrogen column density, associated with 
flaring events, have been observed in the past on \object{Proxima\,Cen} 
(Haisch 
et al. \cite{HLB83}), \object{V773\,Tau} (Tsuboi et al. \cite{TKM98}) and 
\object{Algol} (Ottmann \& Schmitt \cite{OS96}; Favata \& Schmitt 
\cite{FS99}), 
and usually interpreted as due to a coronal mass ejection. However, solar 
observations show that coronal mass ejection and flares are not always 
physically related one to another (Golub \& Pasachoff \cite{GP97}). No 
evidence 
of mass ejection can be significantly singled out from the evolution of the 
best 
fit $N_{\rm H}$ in the present data.  

We modeled the flare using the hydrodynamic decay--sustained heating scenario 
(Reale et al. \cite{RBPSM97}) and assumed that in the 2-T best fit model, the 
hotter temperature represents the flare plasma, while the cooler temperature 
represents the quiescent coronal plasma. We then derived the flare loop 
semi--length that turns out to be quite large, $\sim 1.5\,R_*$. Note that our 
flare temperature (the hotter component) in not well constrained at higher 
values, due to the ROSAT energy band. In any case, since the downward error 
bars 
for the temperature are well constrained by the fit, the loop semi--length 
that 
we derived should be regarded as a lower limit (in the Reale et al. 
\cite{RBPSM97} model scenario). 

We then analyzed the flare with a simple ``order of magnitude" approach 
(Pallavicini et al. \cite{PTS90}, Pallavicini \cite{P95b}) able to provide a 
lower limit for the loop semi--length since it considers only the radiative 
cooling time ignoring the conductive cooling and the continuous heating of the 
flare plasma. The lower limit for the loop semi--length so derived is however 
in 
agreement with that derived by the full hydrodynamic decay--sustained heating 
approach, in spite of the fact that the two techniques are based on quite 
different assumptions. 

We then applied this approach to various flares already analyzed by different 
authors using the Reale et al. (\cite{RBPSM97}) methodology. We considered 
four 
flares from \object{Algol} (Favata et al. \cite{FMRSS00}), three flares from  
\object{AB\,Dor} (Franciosini et al. \cite{Fetal01}), one flare from 
\object{AB\,Dor} (Ortolani et al. \cite{OPMRW98}), six flares from 
\object{AD\,Leo} (Favata et al. \cite{FMR00}) and one flare from 
\object{EV\,Lac} (Favata et al. \cite{FRMSM00}). It is interesting to note 
that 
these simple ``order of magnitude" estimates, based on poorly justified and/or 
physically inconsistent assumptions, give again loop semi--lengths comparable 
to 
those derived with the Reale et al. (\cite{RBPSM97}) model. Apparently, the 
various assumptions made in the simplified ``order of magnitude" approach 
(such 
as for instance that there is no heating during the flare decay or that the 
radiative to conductive losses have a fixed ratio) conspire to give loop sizes 
which are in gross agreement with the method based on detailed hydrodynamic 
calculations. On the other hand, the fact that the latter approach is based on 
physically sound assumptions give a much better confidence on the reliability 
of 
flare loop sizes derived in this way.
 
\begin{acknowledgements}
We want to thank E. Franciosini for having provided us the results of her 
analyses of flares from \object{AB\,Dor} in advance of publication. We
also thank the anonymous referee for her/his suggestions. This work 
has been partially supported by the Italian Space Agency (ASI).
\end{acknowledgements}

\end{document}